\documentclass[prb,twocolumn,superscriptaddress,showpacs,showkeys]{revtex4}
\usepackage[cp1250]{inputenc}
\usepackage{graphics}
\usepackage{amsmath}

\usepackage[%dvips,
pdftitle={Theory of fractional vortex escape in a 0-kappa long Josephson junction},
pdfauthor={Dr. E. Goldobin},
pdfsubject={..},
urlcolor=blue,
bookmarks=true,
bookmarksopen=true,
bookmarksnumbered=true
]{hyperref}

\include{MyMc}

\begin{document}

\title{%
  Theory of fractional vortex escape in a $0$-$\kappa$ long Josephson junction
}

\author{K. Vogel}
\author{W. P. Schleich}
\affiliation{
 Universit\"at Ulm,
 Institut f\"ur Quantenphysik,
 D-89069 Ulm, Germany
}

\author{T. Kato}
\affiliation{
  Institute for Solid State Physics,
  The University of Tokyo,
  Kashiwa, Chiba 277-8581, Japan
}

\author{D.~Koelle}
\author{R.~Kleiner}
\author{E.~Goldobin}
\email{gold@uni-tuebingen.de}
\affiliation{
  Physikalisches Institut -- Experimentalphysik II
  \& Center for Collective Quantum Phenomena,
  Universit\"at T\"ubingen,
  Auf der Morgenstelle 14,
  D-72076 T\"ubingen, Germany
}



\begin{abstract}

We consider a fractional Josephson vortex in an infinitely long 0-$\kappa$ Josephson junction. A uniform bias current applied to the junction exerts a Lorentz force acting on a vortex. When the bias current becomes equal to the critical (or depinning) current, the Lorentz force tears away an integer fluxon and the junction switches to the resistive state. In the presence of thermal and quantum fluctuations this escape process takes place with finite probability already at subcritical values of the bias current. We analyze the escape of a fractional vortex by mapping the Josephson phase dynamics to the dynamics of a single particle in a metastable potential and derive the effective parameters of this potential. This allows us to predict the behavior of the escape rate as a function of the topological charge of the vortex.

\end{abstract}

\date{\today}

\maketitle

\section{Introduction}
\label{Sec:Intro}

Josephson junctions with a phase drop of $\pi$ in the ground state ($\pi$ JJs\cite{Bulaevskii:pi-loop}) are intensively investigated, as they promise important advantages for Josephson junction based electronics\cite{Terzioglu:1997:CompJosLogic,Terzioglu:1998:CJJ-Logic}, and, in particular, for JJ based qubits\cite{Ioffe:1999:sds-waveQubit,Blatter:2001:QubitDesign,Yamashita:2005:pi-qubit:SFS+SIS,Yamashita:2006:pi-qubit:3JJ}. Nowadays several technologies allow to manufacture such junctions: JJs with ferromagnetic barrier\cite{Ryazanov:2001:SFS-PiJJ,Kontos:2002:SIFS-PiJJ,Weides:2006:SIFS-HiJcPiJJ}, quantum dot JJs\cite{vanDam:2006:QuDot:SuperCurrRev,Cleuziou:2006:CNT-SQUID,Jorgensen:2008:QuDotJJ:0-pi-transition} and nonequilibrium superconductor-normal metal-superconductor JJs\cite{Baselmans:1999:SNS-pi-JJ,Huang:2002:NonEquPiJJ}.

%0-pi LJJ, semifluxon
One can also fabricate 0-$\pi$ long Josephson junctions (0-$\pi$ LJJs)\cite{Tsuei:Review,Kirtley:SF:HTSGB,Lombardi:2002:dWaveGB,Smilde:ZigzagPRL,Weides:2006:SIFS-0-pi}, \ie, LJJs some parts of which behave as 0 junctions and other parts as $\pi$ junctions. The ground state phase $\mu(x)$ in such junctions will have the value of 0 deep inside the 0-region, the value of $\pi$ deep inside the $\pi$ region and will continuously change from 0 to $\pi$ in the $\lambda_J$-vicinity of a 0-$\pi$ boundary, where $\lambda_J$ is the Josephson penetration depth. Such a bending of the phase results in the appearance of the magnetic field $\propto d\mu/dx$ localized in the $\lambda_J$-vicinity of a 0-$\pi$ boundary and the supercurrents $\pm\sin[\mu(x)]$ circulating around it, \ie, one deals with a Josephson vortex. The total magnetic flux localized at the 0-$\pi$ boundary is equal to $\Phi_0/2$, where $\Phi_0\approx2.07\times10^{-15}\units{Wb}$ is the magnetic flux quantum. Therefore, such a Josephson vortex is called a semifluxon\cite{Bulaevskii:0-pi-LJJ,Goldobin:SF-Shape,Xu:SF-Shape}. If the Josephson phase $\mu(x)$ deep inside the $\pi$ region is equal to $-\pi$ instead of $\pi$, one will have a localized magnetic flux equal to $-\Phi_0/2$ and a supercurrent of the vortex circulating counterclockwise (antisemifluxon).
Both semifluxons and antisemifluxons were observed experimentally\cite{Hilgenkamp:zigzag:SF} and have been under extensive experimental and theoretical investigation during the last decade.\cite{Kogan:3CrystalVortices,Kirtley:SF:HTSGB,Kirtley:SF:T-dep,Hilgenkamp:zigzag:SF,Kirtley:IcH-PiLJJ,Goldobin:SF-ReArrange,Stefanakis:ZFS/2,Zenchuk:2003:AnalXover,Goldobin:Art-0-pi,Susanto:SF-gamma_c,Goldobin:2KappaGroundStates,Goldobin:F-SF,Kirtley:2005:AFM-SF,Buckenmaier:2007:ExpEigenFreq,Nappi:2007:0-pi:Fiske}

%kappa-vortex
It turns out that instead of a $\pi$-discontinuity of the Josephson phase at 0-$\pi$ boundary one can artificially create any arbitrary $\kappa$-discontinuity of the phase at any point of the LJJ, and the value of $\kappa$ can be tuned electronically\cite{Goldobin:Art-0-pi}. As a result, in the ground state two types of vortices with the topological charges $-\kappa$ and $-\kappa+2\pi$ (we assume that $0<\kappa<2\pi$) can be formed\cite{Goldobin:2KappaGroundStates}. Such vortices are generalizations of semifluxons and antisemifluxons discussed above. They are stable only if their topological charge is smaller than or equal to $2\pi$ by absolute value.

%bias current
When one has a $\kappa$ vortex and applies a spatially uniform bias current through the LJJ, the bias current exerts a Lorentz force, which pushes the vortex along the junction. The direction of the force depends on the mutual polarity of the vortex and the bias current. The vortex exists just because it should compensate the phase discontinuity, and, therefore, it is pinned in the vicinity of the discontinuity, \ie, it may bend under the action of the Lorentz force, but does not move away. Nevertheless, when the Lorentz force becomes strong enough, it tears off a whole integer fluxon out of a $\kappa$ vortex. The fluxon moves away along the junction, while a $\kappa-2\pi$ vortex is left at the discontinuity. Further dynamics leads to the switching of the 0-$\pi$ LJJ into the voltage state. This process was described for the first time for the case of a semifluxon $\kappa=\pi$.\cite{Goldobin:SF-ReArrange} It takes place when the normalized bias current $\gamma=I/I_{c0}$ reaches the critical (depinning) current of\cite{Malomed:2004:ALJJ:Ic(Iinj),Goldobin:F-SF}
\begin{equation}
  \gamma_c(\kappa)=\left| \frac{\sin(\kappa/2)}{\kappa/2} \right|
  , \label{Eq:gamma_c}
\end{equation}
where $I_{c0}=j_c w L$ is the ``intrinsic'' critical current, which corresponds to the measurable critical current if $\kappa=0$; $w$ is the LJJ width and $L$ is its length. 

%fluctuation
In the presence of quantum or thermal fluctuations, the escape process described above will take place with finite probability already at $\gamma<\gamma_c$.
%In this paper
In this paper, we study the escape process of an arbitrary Josephson $\kappa$ vortex by mapping the Josephson phase dynamics to the dynamics of a single particle in an effective one-dimensional metastable potential. This allows us to predict the escape rates as a function of the bias current and of the vortex topological charge in the thermal and quantum domains. Our results can be directly compared with the experimental data that are being obtained currently.

\section{Model}
\label{Sec:Model}

For our calculations we use dimensionless quantities. Lengths are measured in units of the Josephson length $\lambda_J$, times are measured in units of $\omega_p^{-1}$, where $\omega_p$ is the plasma frequency, energies are measured in units of $E_J \lambda_J$ where $E_J$ is the Josephson energy per length, and currents are measured in units of the ``intrinsic'' critical current $I_{c0}$ of the Josephson junction.

The dynamics of a fractional vortex in an infinitely long $0$-$\kappa$
Josephson junction with an applied bias current $\gamma$ is then described by the sine-Gordon equation (dissipation is neglected)
\begin{equation}
\mu_{xx}(x,t) - \mu_{tt}(x,t)
- \sin\left[\mu(x,t) + \kappa H(x) \right] = -\gamma \,,
\label{eq:sine-gordon}
\end{equation}
where $H(x)$ is given by
\begin{equation}
H(x) =
\left\{\begin{array}{ll}
0 & \mbox{for $x<0$} \\[1ex]
1 & \mbox{for $x>0$.}
\end{array}\right.
\end{equation}
This equation can be derived from the Lagrangian density
\begin{equation}
{\cal L}
= \frac{1}{2}\left(\frac{\partial \mu}{\partial t}\right)^2
- \frac{1}{2}\left(\frac{\partial \mu}{\partial x}\right)^2
- U(\mu,x) \,,
\label{eq:L-density}
\end{equation}
where the potential energy density $U(\mu,x)$ is given by
\begin{equation}
  U\left[\mu,x\right] 
  = 1 - \cos\left[\mu+\kappa H(x)\right] -\gamma\mu 
  . \label{eq:U-density}
\end{equation}
The boundary conditions for $\mu(x,t)$ are
\begin{equation}
\mu_x(-\infty,t) = \mu_x(+\infty,t) = 0 \,.
\label{eq:boundary}
\end{equation}
At $x=0$ the Josephson phase $\mu(x,t)$ and its derivative
$\mu_x(x,t)$ are continuous, \ie,
\begin{subequations}
  \begin{eqnarray}
    \mu(0^+,t) = \mu(0^-,t) = \mu(0,t)
    ,\\
    \mu_x(0^+,t) = \mu_x(0^-,t) = \mu_x(0,t)
    .
  \end{eqnarray}
  \label{eq:match}
\end{subequations}

\section{Properties of the stationary solution}
\label{Sec:StatSol}
The stationary solution $\mu_0(x)$ of Eq.~(\ref{eq:sine-gordon}) follows from
\begin{eqnarray}
  \mu_0''(x) - U'\left[\mu_0(x),x\right] =
  \nonumber\\
  \mu_0''(x) - \sin\left[\mu_0(x) + \kappa H(x) \right]+\gamma = 0
  .\label{eq:sine-gordon-stat}
\end{eqnarray}
Without solving this equation we can derive some properties of $\mu_0(x)$ which will be useful for our calculations. Multiplying Eq.~(\ref{eq:sine-gordon-stat}) with $\mu'_0(x)$ in the regions $x<0$ and $x>0$ leads to
\begin{equation}
-\frac{1}{2}\left[\mu'_0(x)\right]^2 + U\left[\mu_0(x),x\right]
= \mathrm{const.}
\label{eq:int}
\end{equation}
Using the boundary conditions~(\ref{eq:boundary}) for the stationary
solution and the abbreviations
\begin{subequations}
  \begin{eqnarray}
    \varphi_- &=& \mu_0(-\infty)
    ;\\
    \varphi_+ &=& \mu_0(+\infty) + \kappa
    ;\\
    \varphi_0 &=& \mu_0(0) + \kappa/2
    , 
  \end{eqnarray}
  \label{eq:varphi_def}
\end{subequations}

we obtain
\begin{widetext}
\begin{eqnarray}
x<0 &:& \frac{1}{2}\,\left[\mu'_0(x)\right]^2
= U\left[\mu_0(x),x\right] - U\left[\mu_0(-\infty),-\infty\right]
= \cos\varphi_{-} - \cos\left[\mu_0(x)\right]
+ \gamma \left[ \varphi_{-} - \mu_0(x)\right]
\nonumber\\[1ex]
x>0 &:& \frac{1}{2}\,\left[\mu'_0(x)\right]^2
= U\left[\mu_0(x),x\right] - U\left[\mu_0(+\infty),+\infty\right]
= \cos\varphi_{+} - \cos\left[\mu_0(x)+\kappa\right]
+ \gamma \left[ \varphi_{+} - \mu_0(x)-\kappa\right] \,.
\label{eq:dphi}
\end{eqnarray}
These two equations allow us to express $\mu_0'(x)$ in terms of $\mu_0(x)$
and $\varphi_{\pm}$ [or $\mu_0(\pm\infty)$]. Furthermore, since $\mu_0(x)$
and $\mu_0'(x)$ are continuous at $x=0$, we have
\begin{equation}
U\left[\mu_0(0),0^-\right] - U\left[\mu_0(-\infty),-\infty\right]
= U\left[\mu_0(0),0^+\right] - U\left[\mu_0(+\infty),+\infty\right]
\end{equation}
which we can rewrite in the form
\begin{equation}
\cos\varphi_{+} - \cos\varphi_{-}
+ \gamma \left(\varphi_{+}-\varphi_{-}-\kappa\right)
= \cos\left[\varphi_0+\kappa/2\right] - \cos\left[\varphi_0-\kappa/2\right]
= -2 \sin\frac{\kappa}{2}\,\sin\varphi_0 \,.
\label{eq:match2}
\end{equation}

For $x\to\pm\infty$ the phase $\mu_0(x)$ approaches a constant value, which minimizes (stable stationary solution) or maximizes (unstable stationary solution) the potential energy density $U\left[\mu_0(x),x\right]$, Eq.~(\ref{eq:U-density}). For stable stationary solutions we find
\begin{equation}
\sin\varphi_{\pm} = \gamma\,,\quad \cos\varphi_{\pm} > 0
\quad\Rightarrow\quad \varphi_{\pm}
   = \arcsin\gamma + 2n_{\pm}\pi\,,\quad
\cos\varphi_{\pm} = \sqrt{1-\gamma^2}\,,
\label{eq:stat_stable}
\end{equation}
whereas for unstable stationary solutions we find
\begin{equation}
\sin\varphi_{\pm} = \gamma\,,\quad \cos\varphi_{\pm} < 0
\quad\Rightarrow\quad \varphi_{\pm}
   = \pi-\arcsin\gamma + 2n_{\pm}\pi\,,\quad
\cos\varphi_{\pm} = -\sqrt{1-\gamma^2}\,.
\label{eq:stat_unstable}
\end{equation}
\end{widetext}
Using these results, Eq.~(\ref{eq:match2}) can be reduced to
\begin{equation}
\gamma \left(2n\pi-\kappa\right)
= -2 \sin\frac{\kappa}{2}\,\sin\varphi_0 \,,
\label{eq:match3}
\end{equation}
where $n=n_{+} - n_{-}$ is the number of fluxons already present in the system. Obviously, this condition cannot be fulfilled for arbitrary values of $\gamma$. At a critical value $\gamma_c$ we will find $\sin\varphi_0 =\pm 1$. For larger values of $\gamma$, Eq.~(\ref{eq:match3}) has no solution for $\varphi_0$, and the stationary solution $\mu_0(x)$ cannot exist. For $n=0$, \ie, $\varphi_{+} = \varphi_{-}$, we obtain the critical current given in Eq.~(\ref{Eq:gamma_c}).

\section{Eigenmodes}
\label{Sec:Eigenmodes}

The stability of the stationary solution $\mu_0(x)$ can be analyzed with the help of the eigenmodes of the sine-Gordon equation~(\ref{eq:sine-gordon}). To find these eigenmodes we insert the ansatz
\begin{equation}
\mu(x,t) = \mu_0(x) + \psi(x)\,e^{-i\omega t}
\end{equation}
into the sine-Gordon equation~(\ref{eq:sine-gordon}) and linearize it, assuming $|\psi(x)| \ll 1$. Since $\mu_0(x)$ solves the stationary sine-Gordon equation, we obtain
the Schr\"odinger equation
\begin{eqnarray}
  &&- \psi''_n(x) + U''\left[\mu_0(x),x\right]\psi_n(x) =
  \nonumber\\
  &=& - \psi''_n(x) + \cos[\mu_0(x) + \kappa H(x)]\psi_n(x) =
  \nonumber\\
  &=& \omega_n^2 \psi_n(x)
  \label{eq:schroedinger}
\end{eqnarray}
for the eigenmodes $\psi_n(x)$, where the index $n$ enumerates the eigenmodes. In this Schr\"odinger equation, the potential is determined by the stationary solution $\mu_0(x)$. Since the boundary conditions for $\mu(x,t)$ are already taken  into account by the stationary solution $\mu_0(x)$, the boundary conditions for the eigenmodes $\psi_n(x)$ read
\begin{equation}
  \psi_n'(-\infty) = \psi_n'(+\infty) = 0\,.
  \label{eq:boundary-psi}
\end{equation}
At $x=0$ the phase $\mu(x,t)$ and its derivative $\mu_x(x,t)$ are continuous, see Eq.~(\ref{eq:match}). Since the stationary solution $\mu_0(x)$ and its derivative $\mu'_0(x)$ are continuous at $x=0$, $\psi_n(x)$ and $\psi'_n(x)$ have to be continuous at $x=0$ as well, \ie,
\begin{subequations}
  \begin{eqnarray}
    \psi _n(0^+) &= \psi _n(0^-) =& \psi _n(0) 
    ;\\
    \psi'_n(0^+) &= \psi'_n(0^-) =& \psi'_n(0) 
    .
  \end{eqnarray}
  \label{eq:match-psi}
\end{subequations}

As long as all eigenvalues $\omega_n^2$ of the Schr\"odinger
equation~(\ref{eq:schroedinger}) are positive, the stationary solution
$\mu_0(x)$ is stable. As soon as one eigenvalue $\omega_n^2$ becomes
negative, the stationary solution $\mu_0(x)$ is unstable. Therefore, at
the critical current $\gamma=\gamma_c$ the lowest eigenvalue, denoted by
$\omega_0^2$, becomes zero.

We now briefly show that at $\gamma=\gamma_c$ the lowest eigenmode
$\psi_0(x)$ is the derivative of the stationary solution, \ie,
$\psi_0(x) = C\mu'_0(x)$. By taking the derivative of the stationary
sine-Gordon equation~(\ref{eq:sine-gordon-stat}) in the regions $x<0$
and $x>0$ we find
\begin{equation}
-\frac{d^2}{dx^2}\, \mu'_0(x)
+ \cos\left[\mu_0(x) + \kappa H(x) \right] \mu'_0(x) = 0 \,.
\label{eq:eigenmode0}
\end{equation}
Therefore, we have found the formal solution $\psi_0(x)=C\mu'_0(x)$ of
the Schr\"odinger equation (\ref{eq:schroedinger}) with an eigenvalue
$\omega_0^2=0$.

The boundary conditions~(\ref{eq:boundary-psi}) and the matching
conditions~(\ref{eq:match-psi}) for the eigenmodes introduce
additional conditions for the stationary solution $\mu_0(x)$:
\begin{subequations}
  \begin{eqnarray}
    && \psi_0'(\pm\infty) = C \mu''_0(\pm\infty) = 0
    \Rightarrow \mu''_0(\pm\infty) = 0
    ,\\[1ex]
    && \psi'_0(0^+) = C \mu''_0(0^+) = \psi'_0(0^-) = C\mu''_0(0^-)
    \nonumber\\  
    && \Rightarrow \mu''_0(0^+) = \mu''_0(0^-)
    .
  \end{eqnarray}
\end{subequations}
Using the stationary sine-Gordon equation~(\ref{eq:sine-gordon-stat}) and
the abbreviations defined in Eq.~(\ref{eq:varphi_def}), we can rewrite these
conditions in the form
\begin{subequations}
  \begin{eqnarray}
    &&\sin\varphi_{-} = \sin\varphi_{+} = \gamma
    ,\\
    &&\sin(\varphi_0-\kappa/2) = \sin(\varphi_0 + \kappa/2)
    \nonumber\\
    \Rightarrow
    &&\sin\frac{\kappa}{2}\cos\varphi_0 = 0.
  \end{eqnarray}
  \label{eq:boundary-psi_c}
\end{subequations}
The first condition is fulfilled for any value of $\gamma$ [see Eqs.~(\ref{eq:stat_stable}) and (\ref{eq:stat_unstable})], whereas the second condition is only true for $\gamma=\gamma_c$, where we have $\sin\varphi_0= \pm 1$ and $\cos\varphi_0 = 0$, see the discussion after Eq.~(\ref{eq:match3}). Therefore, $\psi_0(x)=C\mu'_0(x)$ is only an eigenmode for $\gamma=\gamma_c$, see also Ref.~\onlinecite{Kato:1997:QuTunnel0pi0JJ}.

\section{Effective potential}
\label{Sec:EffectivePotential}

The phase $\mu(x,t)$ can be written in the form
\begin{equation}
\mu(x,t) = \mu_0(x) + \sum\limits_{n} q_n(t)\,\psi_n(x) \,.
\end{equation}
By inserting this expansion into the Lagrangian density~(\ref{eq:L-density}) and integrating over $x$ we can derive a Lagrangian for the mode amplitudes $q_n(t)$, which describes the motion of a fictitious particle in many dimensions. For $\gamma$ close to $\gamma_c$ the eigenfrequency $\omega_0$ approaches zero, whereas the other eigenfrequencies remain finite. Around the minimum at $q_n=0$ the potential will be ``flat'' in the direction of $q_0$ and ``steep'' along the other directions. Therefore, we expect that at low energies, a particle trapped in the minimum of the potential will move along $q_0$. Motivated by this simple picture, we only take into account the dynamics of the mode amplitude $q_0(t)$. To simplify the notation we denote the amplitude of the eigenmode $\psi_0(x)$ by $q(t)$.

The Lagrangian for $q(t)$ can be derived by inserting the ansatz
\begin{equation}
\mu(x,t) \approx \mu_0(x) + q(t)\,\psi_0(x)
\end{equation}
into the Lagrangian density~(\ref{eq:L-density}). Our simplified Lagrangian reads
\begin{widetext}
\begin{equation}
L = \int\limits_{-\infty}^{+\infty}\!{\cal L}\,dx
= \frac{1}{2} \dot{q}^2(t)\!\int\limits_{-\infty}^{+\infty}\!
     \psi_0^{\,2}(x)\, dx
  - \frac{1}{2}\!\int\limits_{-\infty}^{+\infty}\!
     \left[\mu'_0(x) + q(t)\,\psi'_0(x)\right]^2 dx
-\int\limits_{-\infty}^{+\infty}\!
U\left[\mu_0(x) + q(t)\,\psi_0(x),x\right] dx \,,
\end{equation}
where ${\cal L}$ and $U$ are defined in Eqs.~(\ref{eq:L-density})
and (\ref{eq:U-density}). Since we want to describe the escape of a
particle from a metastable potential we only take in to account terms
up to third order in $q(t)$ and use the approximation
\begin{eqnarray}
&& U\left[\mu_0(x) + q(t)\,\psi_0(x),x\right] \approx
U\left[\mu_0(x),x\right]
+ q(t)\,U'\left[\mu_0(x),x\right]\psi_0(x)
\nonumber
\\[1ex]
&& + \frac{1}{2}\,q^2(t)\,U''\left[\mu_0(x),x\right]\psi_0^2(x)
+ \frac{1}{6}\,q^3(t)\,U'''\left[\mu_0(x),x\right]\psi_0^3(x) \,.
\end{eqnarray}
After omitting a constant term we obtain
\begin{eqnarray}
L &=& \frac{1}{2}\,\dot{q}^2(t)
     \!\int\limits_{-\infty}^{+\infty}\! \psi_0^{\,2}(x)\, dx
  - q(t) \!\int\limits_{-\infty}^{+\infty}\!
     \Big\{\mu'_0(x)\,\mu'_0(x)
            +U'\left[\mu_0(x),x\right]\psi_0(x)\Big\}\,dx
\nonumber\\[1ex]
&-& \frac{1}{2}\,q^2(t)\!\int\limits_{-\infty}^{+\infty}\!
\big\{\left[\psi'_0(x)\right]^2
+ U''\left[\mu_0(x),x\right]\psi_0^2(x) \Big\}\,dx
- \frac{1}{6}\,q^3(t)\int\limits_{-\infty}^{+\infty}
U'''\left[\mu_0(x),x\right]\psi_0^3(x)\,dx \,.
\end{eqnarray}

Performing two partial integrations and taking into account the boundary conditions for $\mu_0(x)$ and $\psi_0(x)$ leads to
\begin{eqnarray}
L &=& \frac{1}{2}\,\dot{q}^2(t)
     \!\int\limits_{-\infty}^{+\infty}\!\psi_0^{\,2}(x)\, dx
  - q(t) \!\int\limits_{-\infty}^{+\infty}\!
     \big\{-\mu''_0(x) + U'\left[\mu_0(x),x\right]\big\}
     \psi_0(x)\,dx
\nonumber\\[1ex]
&-&\frac{1}{2}\,q^2(t)\!\int\limits_{-\infty}^{+\infty}\!
   \big\{-\psi''_0(x) + U''\left[\mu_0(x),x\right]\psi_0(x)
   \big\}\psi_0(x)\,dx
- \frac{1}{6}\,q^3(t)\!\int\limits_{-\infty}^{+\infty}\!
U'''\left[\mu_0(x),x\right]\psi_0^3(x) dx \,.
\label{eq:L1}
\end{eqnarray}
The linear term vanishes since $\mu_0(x)$ is the stationary solution
of the stationary sine-Gordon equation~(\ref{eq:sine-gordon-stat}).
In the quadratic term we can use
$-\psi''_0(x) + U''\left[\mu_0(x),x\right]\psi_0(x)=\omega_0^2\psi_0(x)$,
see Eq.~(\ref{eq:schroedinger}). Therefore, the Lagrangian can be written
in the form
\begin{equation}
  L = \frac{1}{2}\,M\,\dot{q}^2(t)
  - \frac{1}{2}M\,\omega_0^2\,q^2(t)
  - \frac{1}{6}G\,q^3(t)
  ,%\label{Eq:Lagrangian}
\end{equation}
where $M$ and $G$ are given by
\begin{equation}
M = \int\limits_{-\infty}^{+\infty}\!\psi_0^2(x)\,dx\,,\qquad
G = \int\limits_{-\infty}^{+\infty}\!  U'''\left[\mu_0(x),x\right]\psi_0^3(x)\,dx
= -\int\limits_{-\infty}^{+\infty}\!
   \sin\left[\mu_0(x)+\kappa H(x)\right] \psi_0^3(x)\,dx\,.
\end{equation}
\end{widetext}

This Lagrangian describes the motion of a fictitious particle of mass $M$
in the effective potential
\begin{equation}
  V_\mathrm{eff}(q) = \frac{1}{2}M \omega_0^2 q^2 + \frac{1}{6}G q^3
  ,\label{eq:Veff}
\end{equation}
which can be characterized by the frequency $\omega_0$ for small oscillations
around the minimum at $q=0$ and a barrier height
\begin{equation}
\Delta V = \frac{2M^3 \omega_0^6}{3G^2} \,.
\label{eq:barrier}
\end{equation}
These two parameters determine the escape rates in the classical and in the
quantum regime, see Sec.~\ref{Sec:Escape}. Note, that the frequency for small
oscillations around the minimum and the eigenfrequency of the lowest eigenmode,
calculated from Eq.~(\ref{eq:schroedinger}), are the same.

The calculation of the present section leads to the following procedure
to determine the frequency $\omega_0$ and the barrier height $\Delta V$:
\begin{enumerate}

\item
For a given bias current $\gamma < \gamma_c$ we solve the stationary
sine-Gordon equation~(\ref{eq:sine-gordon-stat}) numerically and find
the stationary solution $\mu_0(x)$.

\item
For this stationary solution $\mu_0(x)$ we solve the Schr\"odinger
equation~(\ref{eq:schroedinger}) numerically and find the eigenmode
$\psi_0(x)$ and the corresponding eigenvalue $\omega_0^2$.

\item
Using the eigenmode $\psi_0(x)$ we calculate $M$ and $G$ numerically
to find the barrier height $\Delta V$.

\end{enumerate}

\section{Analytical approximations}
\label{Sec:Approx}

\subsection{Approximations for $M$ and $G$}
\label{Sec:Approx_MG}

The approximations for $M$ and $G$ for $\gamma$ close to $\gamma_c$
are straightforward. Since $M$ and $G$ remain finite at $\gamma = \gamma_c$
we replace $M$ and $G$ by their values at $\gamma = \gamma_c$ and use
the not normalized eigenmode $\psi_0(x)=\mu_0'(x)$, \ie,
\begin{subequations}
  \begin{eqnarray}
    M &\approx& \int\limits_{-\infty}^{+\infty}
    \left[\mu_0'(x)\right]^2 dx
    ,\\
    G &\approx& - \int\limits_{-\infty}^{+\infty}
    \sin\left[\mu_0(x) + \kappa H(x)\right] \left[\mu_0'(x)\right]^3 dx
    .
  \end{eqnarray}
\end{subequations}

These two integrals can be calculated analytically to a large extent. As shown in Appendix~\ref{Sec:App-MG}, the expressions for $M$ and $G$ can be written in the form
\begin{widetext}
\begin{equation}
M =
\pm\sqrt{2}\!\!\!\int\limits_{\varphi_{-}}^{\varphi_0-\kappa/2}\!\!\!
\sqrt{\cos\varphi_{-} -\cos\varphi + \gamma_c\left(\varphi_{-}
- \varphi\right)}\;\;d\varphi
\pm\sqrt{2}\!\!\!\int\limits_{\varphi_0+\kappa/2}^{\varphi_{+}}\!\!\!
\sqrt{\cos\varphi_{+} -\cos\varphi + \gamma_c\left(\varphi_{+}
- \varphi\right)}\;\;d\varphi
\label{eq:M_approx}
\end{equation}
and
\begin{equation}
G = 2 \sin\frac{\kappa}{2}\,\sin\varphi_0
\Big[
   \cos\varphi_{+} + \cos\varphi_{-}
 + \gamma_c \left(\varphi_{+}+\varphi_{-}-2\varphi_0\right)
\Big] \,,
\label{eq:G_approx}
\end{equation}
\end{widetext}
where $\varphi_{+}$, $\varphi_{-}$, and $\varphi_0$ follow from the behavior of the stationary solution for $\gamma=\gamma_c$ at $x=\pm\infty$ and $x=0$, see Eq.~(\ref{eq:varphi_def}). The upper sign for $M$ applies to vortices with $\mu'_0(x) > 0$, whereas the lower sign applies to vortices with $\mu'_0(x) < 0$.

For our examples in Sec.~\ref{Sec:Results} we use vortices with
\begin{subequations}
  \begin{eqnarray}
    \mu_0(-\infty) &=& \arcsin\gamma
    ,\\
    \mu_0(+\infty) &=& \arcsin\gamma - \kappa
  \end{eqnarray}
\end{subequations}
to satisfy Eq.~(\ref{eq:stat_stable}). In the limit $\gamma\to\gamma_c$ we therefore have $\varphi_{+} = \varphi_{-} = \arcsin\gamma_c = \varphi_c$ which for positive bias currents implies $\varphi_0 = \pi/2$. Moreover, $\mu'_0(x)$ is positive for $-2\pi<\kappa<0$ and negative for $0<\kappa<2\pi$.  In this case the expressions \eqref{eq:M_approx} and \eqref{eq:G_approx} for $M$ and $G$ reduce to 
\begin{subequations}
  \begin{eqnarray}
    M &=& \sqrt{2}\!\!\!\!\!\!\!
    \int\limits_{(\pi-|\kappa|)/2}^{(\pi+|\kappa|)/2}
    \!\!\!\!\!\!\!
    \sqrt{
      \cos\varphi_c -\cos\varphi + \gamma_c\left(\varphi_c-\varphi\right)
    }d\varphi 
    ,\\
    G &=& 4\sin\frac{\kappa}{2} \Big[
      \cos\varphi_c + \gamma_c \left(\varphi_c - \pi/2\right)
    \Big]
    .
  \end{eqnarray}
\end{subequations}

\subsection{Approximations for $\omega_0$ and $\Delta V$}
\label{Sec:Approx_omega}

Since $\omega_0^2$ vanishes for $\gamma\to\gamma_c$ we cannot use its value at $\gamma_c$. Instead we use the stationary solution $\mu_0(x)$ and the eigenmode $\psi_0(x)=\mu_0'(x)$ at $\gamma=\gamma_c$ to evaluate the Lagrangian~(\ref{eq:L1}) for $\gamma<\gamma_c$. In this case, the quadratic term in Eq.~(\ref{eq:L1}) vanishes since $U''[\mu_0(x),x]$ does not depend on $\gamma$. The linear term, however, does not vanish since $\mu_0(x)$ is not the stationary solution for $\gamma<\gamma_c$. Here we have 
\begin{eqnarray}
  &&-\mu''_0(x) + U'\left[\mu_0(x),x\right]
  \nonumber\\
  &=& -\mu''_0(x) + \sin\left[\mu_0(x) + \kappa H(x)\right] -\gamma
  = \gamma_c -\gamma
  .
\end{eqnarray}
The remaining integral can easily be calculated and the approximate
effective potential reads
\begin{equation}
  V_\mathrm{eff}(q) \approx \Delta\mu\,(\gamma_c-\gamma)\,q
  + \frac{1}{6}G\,q^3 \,,
\end{equation}
where $\Delta\mu$ is given by
\begin{eqnarray}
  \Delta\mu &=& \int\limits_{-\infty}^{+\infty}\psi_0(x)\,dx
  = \int\limits_{-\infty}^{+\infty} \mu'_0(x)\,dx
  \nonumber\\
  &=& \mu_0(+\infty) - \mu_0(-\infty) 
  = \varphi_{+} - \varphi_{-} -\kappa
  .
\end{eqnarray}

For this potential we find the frequency
\begin{equation}
\omega_0^{\mathrm{cr}} = \left[
  \frac{2G\Delta\mu}{M^2}\,(\gamma-\gamma_c)
\right]^{1/4}
\label{Eq:omega_cr}
\end{equation}
for small oscillations around the minimum and the barrier height
\begin{equation}
\Delta V^{\mathrm{cr}} = \frac{2}{3}|G|
\left[\frac{2\Delta\mu}{G}\,(\gamma-\gamma_c)\right]^{3/2}
= \frac{2M^3 [\omega_0^{\mathrm{cr}}]^6}{3G^2} \,.
\label{Eq:DeltaV_cr}
\end{equation}
The first expression for $\Delta V^{\mathrm{cr}}$ does not depend on $M$
whereas the second expression for $\Delta V^{\mathrm{cr}}$ agrees with
Eq.~(\ref{eq:barrier}), except that now the approximate expressions for $M$,
$G$, and $\omega_0$ are used.

\subsection{Point-like Josephson junction}
\label{Sec:JJ-point}

We would like to compare our results to the point-like JJ in the following sense. At given $\kappa$, in experiments one sees a certain value of $I_c$ and can calculate expected eigenfrequency, barrier height and escape rates using a short JJ model which ignores phase discontinuities and details related to the internal structure of the solution. We compare these results with our results for a fractional vortex. To obtain the results from the point-like JJ model we insert the values of $\gamma$ and $\gamma_c$ for a long $0$-$\kappa$ JJ into the well known expressions for a point-like JJ\cite{Wallraff:2003:LJJ:ThermEsc}   and obtain
\begin{equation}
  \Delta V^{\mathrm{pt}} = 2l \left[
     \sqrt{1-\left(\frac{\gamma}{2\gamma_c}\right)^2}
     - \frac{\gamma}{\gamma_c}\arccos\,\frac{\gamma}{\gamma_c}
  \right]
  \label{Eq:DeltaV_pt}
\end{equation}
and
\begin{equation}
  \omega_0^{\mathrm{pt}} = \sqrt{\gamma_c}
  \left[1-\left(\frac{\gamma}{\gamma_c}\right)^2\right]^{1/4}
  ,\label{Eq:omega_pt}
\end{equation}
where $l=L/\lambda_J$ is the normalized length of the Josephson junction.

\section{Escape rates}
\label{Sec:Escape}

In Sec.~\ref{Sec:EffectivePotential} we have found that we can map the escape of a fractional vortex to the escape of a particle in a one-dimensional metastable potential which is characterized by the frequency $\omega_0$ for small oscillations around the minimum and a barrier height $\Delta V$. For such
potentials approximate expressions for escape rates are available in the
literature. As we use scaled quantities in the present paper we introduce the
scaled temperature $\theta$ which measures temperatures in units of
$E_J \lambda_J/k_B$ and the dimensionless parameter
$\eta = \hbar\omega_p/(E_J\lambda_J)$ which plays the role of an effective
$\hbar$.

In the classical regime, the escape of a particle from a metastable potential
is due to thermal hopping. The escape rate (measured in units of $\omega_p$)
for such processes is given by Kramers' formula\cite{Haenggi:1990:Kramers50,Weiss-1999}
\begin{equation}
\Gamma_\mathrm{th} = \rho\,\frac{\omega_0}{2\pi} e^{-\Delta V/\theta} ,
\label{eq:Kramers}
\end{equation}
where the prefactor $\rho$ depends on the damping constant $\alpha$
of the system. For $\alpha/\omega_0 \gtrsim 5 \theta/(36 \Delta V)$ it reads
\begin{equation}
\rho = \sqrt{1 + \left(\frac{\alpha}{2\omega_0}\right)^2}
       - \frac{\alpha}{2\omega_0} \,.
\label{eq:rho}
\end{equation}
If the damping constant $\alpha$ becomes too small, it has to be replaced by
\begin{equation}
\rho = \frac{36\alpha\Delta V}{5\theta\omega_0} \,.
\end{equation}

In the quantum regime, the escape of a particle from a metastable
potential is due to tunneling trough the energy barrier. In the
semiclassical limit the decay rate of the ground state of a cubic metastable
potential is given by\cite{Coleman:1977:Semiclassics,Caldeira:1983:DissQT,Weiss-1999}
\begin{equation}
\Gamma_\mathrm{qm} = \sqrt{60}\;\omega_0\;\sqrt{\frac{18 \Delta V}{5\pi\eta\omega_0}}\;
\exp\left(-\frac{36}{5}\,\frac{\Delta V}{\eta\omega_0}\right) \,.
\label{Eq:QuRate:Caldeira-Leggett}
\end{equation}

According to Ref.~\onlinecite{Weiss-1999} for Ohmic damping the crossover
between thermal hopping and quantum tunneling occurs at the temperature
\begin{equation}
  \theta^{\star} = \rho\,\frac{\eta\omega_0}{2\pi}
  , \label{eq:Tcrossover}
\end{equation}
where $\rho$ is given by Eq.~(\ref{eq:rho}).

Before we present our results we have a closer look at Eqs.~(\ref{eq:Kramers})
and (\ref{Eq:QuRate:Caldeira-Leggett}), in particular for small damping where
we can use $\rho \approx 1$. For both equations, the escape rates should be
exponentially small, \ie, the exponents $\Delta V/\theta$ and
$36 \Delta V/(5\eta\omega_0)$ should be large. Using Eq.~(\ref{eq:barrier}) we
find that Kramers' formula, Eq.~(\ref{eq:Kramers}), can be applied if
$\omega_0$ satisfies
\begin{equation}
  \omega_0^6 \gg \frac{3\,G^2}{2\,M^3}\,\theta\,,\quad
  \omega_0^5 \gtrsim \frac{5\,G^2}{24\,M^3}\,\frac{\theta}{\alpha}\,,\quad
  \omega_0  \gg \alpha/2
  .\label{Eq:Kramers:valid}
\end{equation}
The last two conditions allow us to use $\rho \approx 1$. The semiclassical expression for the quantum mechanical escape rate~(\ref{Eq:QuRate:Caldeira-Leggett}) can be used for
\begin{equation}
  \omega_0^5 \gg \frac{5\,G^2}{24\,M^3}\,\eta \,,\quad
  \omega_0 \gg \theta/\eta
  .\label{Eq:Semiclassic:valid}
\end{equation}
If the second condition is violated, the system will not be in the quantum mechanical ground state, and we have to take into account the quantum decay from exited states which is not included in Eq.~(\ref{Eq:QuRate:Caldeira-Leggett}). For $\eta\omega_0 \ll \Delta V$ (many ``bound'' states in the metastable potential) we may thermally average the decay rates from the ground state and the excited states. For details see Ref.~\onlinecite{Weiss-1999}.

The conditions introduced so far define lower bounds for $\omega_0$. On the other hand our simple model of a particle moving in a one-dimensional potential is only valid for $\gamma$ close to $\gamma_c$ where $\omega_0$ becomes small. The main assumption we used to map the full problem to a one-dimensional problem was that $\omega_0^2$ is much smaller than the other eigenvalues of the Schr\"odinger equation~(\ref{eq:schroedinger}). Therefore, we have to require $\omega_0^2 \ll \omega_1^2$, where $\omega_1^2$ is the eigenvalue of the first excited state of the Schr\"odinger equation~(\ref{eq:schroedinger}), or the edge of the plasma band.

\section{Results}
\label{Sec:Results}

Our numerical results are based on Eqs.~(\ref{eq:sine-gordon-stat}) and (\ref{eq:schroedinger}). We solve these equations numerically for a symmetric junction with a length of $20\lambda_J$ to emulate an infinitely long JJ and use vortices with $\mu_0(-\infty) = \mu_0(\infty)+\kappa = \arcsin\gamma$ to satisfy Eq.~(\ref{eq:stat_stable}). Two additional parameters are necessary to calculate escape rates: $E_J\lambda_J/k_B$ is the conversion factor between the scaled temperature $\theta$ and the temperature $T$, and $\eta=\hbar\omega_p/(E_J\lambda_J)$ plays the role of an effective $\hbar$. 

For our examples we use the following JJ parameters: critical current density $j_c=100\units{A/cm^2}$, specific capacitance $C= 4.2 \units{\mu F/cm^2}$, junction width $w=1\units{\mu m}$. This gives $\omega_p = 2\pi\cdot 42.8\units{GHz}$ and $E_J\lambda_J = 78.4\units{meV}=909\units{K}$.  For these parameters, $T$ and $\theta$ are related via $T = E_J \lambda_J/k_B \cdot\theta \approx 909 \units{K}\cdot\theta$, and the value of $\eta$ is $\eta=2.3\cdot 10^{-3}$. Furthermore, we assume that we are in the semiclassical limit, where we can apply Eq.~(\ref{Eq:QuRate:Caldeira-Leggett}) and that we can use $\rho \approx 1$ in Eqs.~(\ref{eq:Kramers}) and (\ref{eq:Tcrossover}). 

%%% assumption or sure?

\subsection{Comparison of different methods}

Before we present our results for escape rates, we first compare the approximations for $\omega_0$ and $\Delta V$ presented in Sec.~\ref{Sec:Approx} to the corresponding numerical values based on the single-mode approximation of Sec.~\ref{Sec:EffectivePotential}. For this purpose we calculate the eigenfrequency and energy barrier using three approaches: (a) single-mode approximation for $\gamma<\gamma_c$, Eq.~(\ref{eq:schroedinger}) for $n=0$ and Eq.~(\ref{eq:barrier}), denoted with superscript ``nu''; (b) approximation at $\gamma=\gamma_c$, Eqs.~(\ref{Eq:omega_cr}) and (\ref{Eq:DeltaV_cr}), denoted with superscript ``cr''; and (c) point-like JJ formulas~(\ref{Eq:DeltaV_pt}) and (\ref{Eq:omega_pt}), denoted with superscript ``pt''.

\begin{figure}[!t]
  \includegraphics{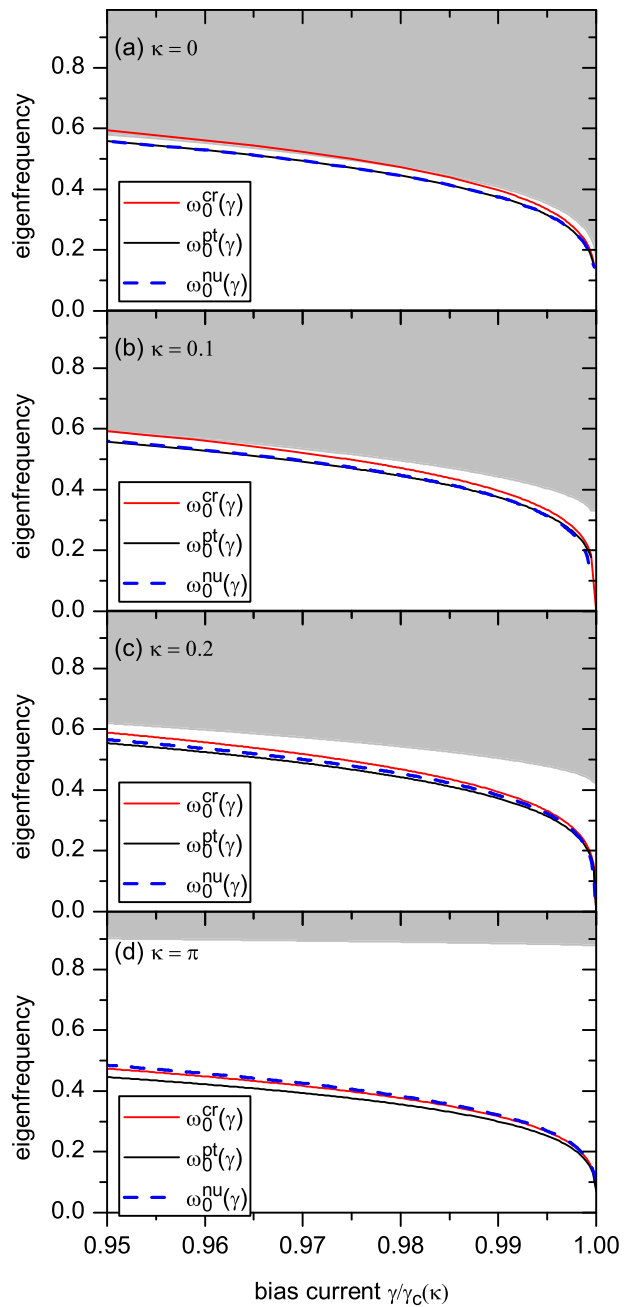}
  \caption{(Color online)
    The frequencies $\omega_0^\mathrm{cr}$, $\omega_0^\mathrm{pt}$ and $\omega_0^\mathrm{nu}$ as a function of the bias current $\gamma$ for different values of $\kappa$. The gray area indicates the plasma band, which was drawn by filling up the area above $\omega_1^{\mathrm{nu}}(\gamma)$.
  }
  \label{Fig:Cmp:Omega(gamma)@diff.kappa}
\end{figure}

The eigenfrequencies are shown in Fig.~\ref{Fig:Cmp:Omega(gamma)@diff.kappa}. At $\kappa=0$ the eigenfrequency $\omega_0^\mathrm{nu}$ coincides with $\omega_0^\mathrm{pt}$, while $\omega_0^\mathrm{cr}$ provides a very reasonable approximation. At larger values of $\kappa$ both $\omega_0^\mathrm{pt}$ and $\omega_0^\mathrm{cr}$ provide good approximations to $\omega_0^\mathrm{nu}$, but underestimate it a little.

For our single-mode approximation to be valid, we have to make sure that the higher eigenfrequencies $\omega_n$ are much larger than $\omega_0$. From Fig.~\ref{Fig:Cmp:Omega(gamma)@diff.kappa} one can see that this is not the case for small values of $\kappa$ where the eigenfrequency of a fractional vortex is close to the edge of the plasma band (shown in gray). Therefore, the \emph{single-mode approximation fails to describe the escape process in a long JJ without discontinuities}. On the other hand, Fig.~\ref{Fig:Cmp:Omega(gamma)@diff.kappa} shows the plasma band for an infinitely long JJ. For a JJ of finite normalized length $l$ the plasma band consists of a set of discrete frequencies $\omega_n$, where the spacing between $\omega_n$ is roughly inversely proportional to $l$. For moderate and especially for small values of $l$ the difference between $\omega_0$ and the other eigenfrequency becomes large, and the single-mode approximation works again even for $\kappa \to 0$ (point-like JJ formula).
\begin{figure}[!t]
  \includegraphics{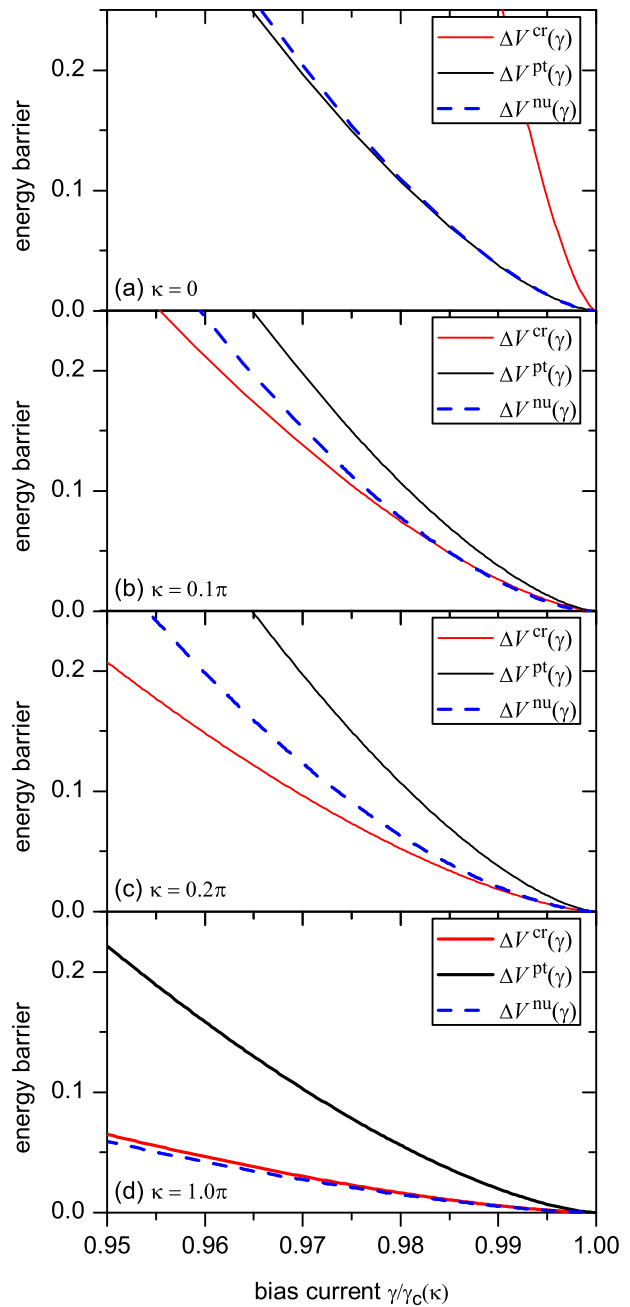}
  \caption{(Color online)
    The energy barriers $\Delta V^\mathrm{cr}$, $\Delta U^\mathrm{pt}$ and $\Delta U^\mathrm{nu}$ as a function of the bias current $\gamma$ for different values of $\kappa$.
  }
  \label{Fig:Cmp:Barrier(gamma)@diff.kappa}
\end{figure}

The energy barrier calculated using different methods is shown in Fig.~\ref{Fig:Cmp:Barrier(gamma)@diff.kappa}. For small values of $\kappa$, $\Delta V^\mathrm{pt}$ provides an excellent approximation to $\Delta V^\mathrm{nu}$ as expected in this limit, while $\Delta V^\mathrm{cr}$ quite overestimates the barrier. For large values of $\kappa$ ($\kappa=\pi$) the situation reverses: $\Delta V^\mathrm{cr}$ approximates $\Delta V^\mathrm{nu}$ very well, while $\Delta V^\mathrm{pt}$ gives overestimated values. The latter is expected as it is easier to activate a bent phase string starting the activation from some point than to move a flat string simultaneously over the barrier. It seems that starting from $\kappa\sim0.2\pi$ the values of $\Delta V^\mathrm{cr}$ provide good approximations to $\Delta V^\mathrm{nu}$. Note, that we have chosen $l=20$ to emulate an infinitely long JJ as the vortex solution is localized on the length scale $1\ll l$. However, for $\kappa\to0$ the phase becomes flat and looses its localization, so that both $\Delta V^\mathrm{nu}$ and $\Delta V^\mathrm{pt}$ become $\propto l$, whereas  $\Delta V^\mathrm{cr}$ does not depend on $l$. In any case, for $\kappa\to0$ the single (lowest) mode approximation does not work, so that one does not have to worry about the discrepancies in $\Delta V$ in this limit.

\subsection{Quantum tunneling}

\begin{figure}[!tb]
  \includegraphics{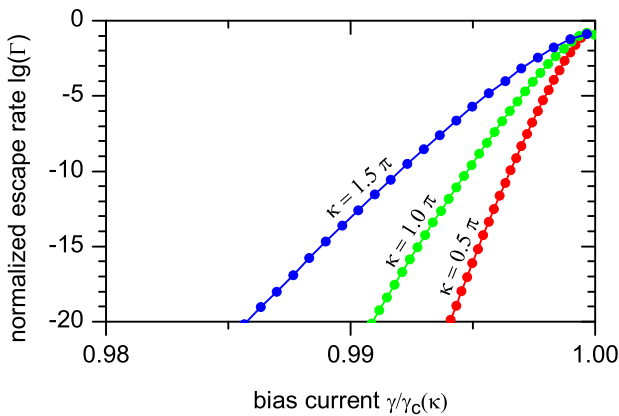}
  \caption{(Color online)
    Quantum escape rates $\Gamma_\mathrm{qm}$ as a function of the bias current $\gamma$ for different values of $\kappa$.
  }
  \label{Fig:QuEscRates@kappa}
\end{figure}

We use the numerically calculated values for the eigenfrequency $\omega_0^\mathrm{nu}$ and barrier height $\Delta V^\mathrm{nu}$ and apply the semiclassical formula~\eqref{Eq:QuRate:Caldeira-Leggett} to these values to calculate quantum tunneling rates. The results are shown in Fig.~\ref{Fig:QuEscRates@kappa}. It turns out that the escape rate is larger, \ie, the vortex escapes easier, for larger values of $\kappa$. This result is qualitatively understandable because for $\kappa\to0$ the escape process reminds more and more the escape of the flat string from the metastable minimum and should scale with the JJ length.

\subsection{Thermal vs. quantum escape}

\begin{figure}[!tb]
  \includegraphics{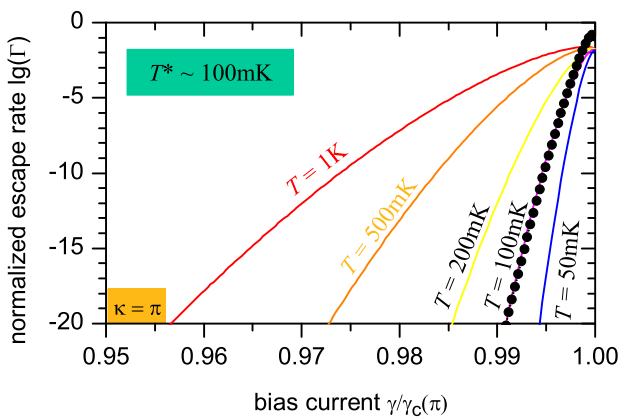}
  \caption{(Color online)
   Thermal escape rates $\Gamma_\mathrm{th}$ as a function of the bias current $\gamma$ for different temperatures. Dots represent the quantum escape rate $\Gamma_\mathrm{qm}$. All escape rates are calculated for $\kappa=\pi$.
  }
  \label{Fig:ThVsQu}
\end{figure}

We use the numerically calculated values for the eigenfrequency $\omega_0^\mathrm{nu}$ and barrier height $\Delta V^\mathrm{nu}$ and apply Eqs.~(\ref{eq:Kramers}) with $\rho=1$ and (\ref{Eq:QuRate:Caldeira-Leggett}) to calculate escape rates at different temperatures $T$, as shown in Fig.~\ref{Fig:ThVsQu}for $\kappa=\pi$. The dots represent the quantum escape rate. We remind that these rates were obtained using the Eqs.~(\ref{eq:Kramers}) and (\ref{Eq:QuRate:Caldeira-Leggett}) that are not valid very close to $\gamma_c(\kappa)$, since in this case the escape rates are not exponentially small. Estimations using Eqs.~\eqref{Eq:Kramers:valid} and \eqref{Eq:Semiclassic:valid} show that expressions \eqref{eq:Kramers} at $T=100\units{mK}$ and expression \eqref{Eq:QuRate:Caldeira-Leggett} become invalid for $\gamma>0.999\gamma_c(\pi)$. One can see that for $T = 100\units{mK}$ the two escape rates agree very well for $\gamma$ not very close to 1. We may therefore define a crossover temperature $T^{\star}\approx 100\units{mK}$ --- independent of $\gamma$ in the range of validity of \eqref{eq:Kramers} and \eqref{Eq:QuRate:Caldeira-Leggett}.

On the other hand, for the parameters used above Eq.~(\ref{eq:Tcrossover}) with $\rho=1$ and $\gamma=0$ gives $T^\star\approx330 \units{mK}$. 
According to Fig.~\ref{Fig:Cmp:Omega(gamma)@diff.kappa}(d) in the region of interest the eigenfrequency $\omega_0(\pi,\gamma)$ is about 2 times smaller that $\omega(\kappa,0)$. Thus Eq.~(\ref{eq:Tcrossover}) predicts $T^* \approx 160 \units{mK}$ in this region.

\section{Conclusions}
\label{Sec:Conclusions}

We have investigated the thermal and quantum escape of an arbitrary fractional Josephson vortex close to its depinning current in an infinitely long Josephson junction. By using a single (lowest) mode approximation, we have mapped the dynamics of an infinite dimensional system to the problem of a point-like particle escaping from a 1D metastable cubic potential. For vanishing topological charge, the single mode approximation fails because the lowest eigenmode is not well separated from the rest of the excitation spectrum. Thus, the lowest mode approximation cannot be used to describe the escape in a conventional long JJ ($\kappa=0$). 

In the region of validity of the single mode approximation we have calculated the eigenfrequency and the barrier height numerically and analytically close to the depinning current. Then we have used the Kramers' formula and a semiclassical expression for thermal and quantum escape rates, respectively, to compare the escape rates of vortices with different topological charges and find the thermal-to-quantum crossover temperature. We have found that vortices with a larger topological charge escape easier. For typical experimental parameters the crossover temperature lays in the range of $100\units{mK}$ as for many other JJ systems. These results can be directly compared to experiments that are in progress in the T\"ubingen group. 

\begin{acknowledgments}

Financial support by the DFG (project SFB/TRR-21) is gratefully acknowledged.

\end{acknowledgments}

\appendix

\section{Integrals}
\label{Sec:App-MG}

In this appendix we evaluate the integrals for $M$ and $G$. The calculations of this appendix are based on the unnormalized eigenmode $\psi_0(x) = \mu'_0(x)$ which is only valid for $\gamma = \gamma_c$. Furthermore, we use the abbreviations $\varphi_{+}$, $\varphi_{-}$ and $\varphi_0$ as defined in Eq.~(\ref{eq:varphi_def}).

\begin{widetext}
With the help of Eq.~(\ref{eq:dphi}) the integral in the definition
of $G$ can be evaluated analytically for $\gamma=\gamma_c$:
\begin{eqnarray}
G &=& -\int\limits_{-\infty}^{+\infty}
\sin\left[\mu_0(x)+\kappa H(x)\right]
\left[\mu'_0(x)\right]^3 dx
= -\int\limits_{-\infty}^{0}
\sin\left[\mu_0(x)\right] \left[\mu'_0(x)\right]^3 dx
- \int\limits_{0}^{\infty}
\sin\left[\mu_0(x)+\kappa\right] \left[\mu'_0(x)\right]^3 dx
\nonumber\\[1ex]
&=& -2 \!\!\!\int\limits_{\mu_0(-\infty)}^{\mu_0(0)}\!\!\!
\sin\varphi \left[
   \cos\varphi_{-} -\cos\varphi
  + \gamma_c\left(\varphi_{-} - \varphi\right)
\right] d\varphi
\nonumber\\[1ex]
&& - 2 \!\!\!\int\limits_{\mu_0(0)}^{\mu_0(+\infty)}\!\!\!
\sin\left(\varphi+\kappa\right)
\left[
\cos\varphi_{+}
- \cos\left(\varphi + \kappa \right)
+ \gamma_c\left(\varphi_{+} - \varphi - \kappa\right)
\right] d\varphi
\nonumber\\[1ex]
&=& -2 \!\!\!\int\limits_{\varphi_{-}}^{\varphi_0-\kappa/2}\!\!\!
\sin\varphi \left[
\cos\varphi_{-} - \cos\varphi + \gamma_c\left(\varphi_{-}-\varphi\right)
\right]
d\varphi
- 2 \!\!\!\int\limits_{\varphi_0+\kappa/2}^{\varphi_{+}}\!\!\!
\sin\varphi \left[
\cos\varphi_{+} - \cos\varphi + \gamma_c\left(\varphi_{+}-\varphi\right)
\right]
d\varphi
\nonumber\\[1ex]
&=&-2 \Bigg[
  \frac{1}{2}\cos^2\varphi
  - \cos\varphi_{-}\,\cos\varphi
  -\gamma_c\sin\varphi
  -\gamma_c\left(\varphi_{-}-\varphi\right)\cos\varphi
\Bigg]_{\varphi_{-}}^{\varphi_0-\kappa/2}
\nonumber\\[1ex]
&& -2 \Bigg[
  \frac{1}{2}\cos^2\varphi
  - \cos\varphi_{+}\,\cos\varphi
  -\gamma_c\sin\varphi
  - \gamma_c\left(\varphi_{+}-\varphi\right)\cos\varphi
\Bigg]_{\varphi_0+\kappa/2}^{\varphi_{+}}
\nonumber\\[1.5ex]
& = & - \cos^2(\varphi_0-\kappa/2)
      - 2\gamma_c\sin(\varphi_0-\kappa/2)
      + 2 \cos(\varphi_0-\kappa/2) \left[
      \cos\varphi_{-} + \gamma_c (\varphi_{-}-\varphi_0+\kappa/2) \right]
\nonumber\\[1.5ex]
&&    - \cos^2\varphi_{-} + \cos^2\varphi_{+}
      + 2\gamma_c(\sin\varphi_{-}-\sin\varphi_{+})
      + \cos^2(\varphi_0+\kappa/2)
      - 2\gamma_c\sin(\varphi_0+\kappa/2)
\nonumber\\[1.5ex]
&&    - 2 \cos(\varphi_0+\kappa/2) \left[
        \cos\varphi_{+} + \gamma_c (\varphi_{+}-\varphi_0-\kappa/2) \right] \,.
\end{eqnarray}
Using Eqs.~(\ref{eq:match2}) and (\ref{eq:boundary-psi_c})we finally arrive at
\begin{equation}
G = 2 \sin\frac{\kappa}{2}\,\sin\varphi_0
\Big[
   \cos\varphi_{+} + \cos\varphi_{-}
 + \gamma_c \left(\varphi_{+}+\varphi_{-}-2\varphi_0\right)
\Big] \,.
\end{equation}

In a similar way we can derive an expression for $M$. With the help of
Eq.~(\ref{eq:dphi}) we obtain
\begin{eqnarray}
M &=& \int\limits_{-\infty}^{+\infty}\left[\mu'_0(x)\right]^2 dx
= \int\limits_{-\infty}^{0}\left[\mu'_0(x)\right]^2 dx
+ \int\limits_{0}^{+\infty}\left[\mu'_0(x)\right]^2 dx
\nonumber\\[1ex]
&=& \pm\sqrt{2}\!\!\!\int\limits_{\mu_0(-\infty)}^{\mu_0(0)}\!\!\!
\sqrt{
   \cos\varphi_{-} -\cos\varphi
  + \gamma_c\left(\varphi_{-} - \varphi\right)
}\;\;d\varphi
\pm\sqrt{2}\!\!\!\int\limits_{\mu_0(0)}^{\mu_0(+\infty)}\!\!\!
\sqrt{
\cos\varphi_{+}
- \cos\left(\varphi + \kappa \right)
+ \gamma_c\left(\varphi_{+} - \varphi - \kappa\right)
}\;\;d\varphi
\nonumber\\[1ex]
&=& \pm\sqrt{2}\!\!\int\limits_{\varphi_{-}}^{\varphi_0-\kappa/2}\!\!
\sqrt{\cos\varphi_{-} -\cos\varphi + \gamma_c\left(\varphi_{-}
- \varphi\right)}\;\;d\varphi
\pm\sqrt{2}\!\!\int\limits_{\varphi_0+\kappa/2}^{\varphi_{+}}\!\!
\sqrt{\cos\varphi_{+} -\cos\varphi + \gamma_c\left(\varphi_{+}
- \varphi\right)}\;\;d\varphi \,.
\end{eqnarray}
The upper sign applies to vortices with $\mu'_0(x) > 0$
whereas the lower sign applies to vortices with $\mu'_0(x) < 0$.
\end{widetext}

\bibliography{MyJJ,SF,pi,SFS,QuComp,QuMech,JJ}

\end{document}